\documentstyle[12pt]{article}
\begin{document}
\newcommand{\hlam}{\mbox{$H_{\lambda}$}}
\newcommand{\ad}[1]{\mbox{${#1}^{\dagger}$}}
\newcommand{\kone}[1]{\mbox{$|{#1}\rangle$}}
\newcommand{\ktwo}[2]{\mbox{$|{#1},{#2}\rangle$}}
\newcommand{\kthree}[3]{\mbox{$|{#1}\,;\,{#2}_{#3}\rangle$}}
\newcommand{\be}{\begin{eqnarray}}
\newcommand{\ee}{\end{eqnarray}}

\title{Geometric phases for generalised squeezed coherent states}
\author{S. Seshadri, S. Lakshmibala and V. Balakrishnan\\
{\em Department of Physics, Indian Institute of Technology,}\\
{\em Madras 600 036,
India}}
\date{}
\maketitle
\vspace{.1cm}
\begin{abstract}
 A simple technique is used to obtain a general formula for the Berry phase
(and the corresponding Hannay angle) for an arbitrary Hamiltonian with an 
equally-spaced spectrum and appropriate ladder operators connecting the 
eigenstates. The formalism is first applied to a general deformation of the 
oscillator involving both squeezing and displacement. Earlier results are
shown to emerge as special cases. The analysis is then extended to multiphoton 
squeezed coherent states and the corresponding anholonomies deduced.
\end{abstract}

\mbox{}\hspace{.3cm}{\bf PACS Nos}. 03.65.Bz, 42.50 Dv, 03.65 Sq
\newpage

\section{Introduction}
Generalised coherent states of various kinds have been discussed in recent 
years in the literature (see, e.g., [1-7]). These states play an important 
role in multiphoton processes in quantum optics, and also have applications 
in quantum measurement theory. A unified description of
multiphoton coherent states has been given recently [8]. Many of these states 
can be identified with eigenstates of Hamiltonians that are essentially number 
operators in appropriate Fock spaces, so that the corresponding levels are 
equally spaced. While some of these states have classical properties, others 
like the `cat states' [9,10] (eigenstates of $a^m$, where $a$ is the 
annihilation operator) are non-classical. In view of the significance of 
these states in optics, an investigation of their quantum (and wherever 
possible, classical) properties is of interest. 

An important aspect in this regard is the geometric phase or
anholonomy associated with the evolution of these states in
certain circumstances. Originally derived by Berry [11] for
Hamiltonians with a non-degenerate spectrum under a cyclic,
adiabatic variation of parameters, the formalism has been
extended to Hamiltonians with a degenerate spectrum [12] as well
as non-adiabatic [13] and non-cyclic [14] variation of
parameters. Even more general settings for the geometric phase
have been pointed out, involving a group-theoretic approach [15]
and a quantum-kinematic approach [16].

In terms of physical applications, the geometric phase and its
generalizations have attracted a lot of interest in a wide
variety of fields (e.g., see [17,18]), especially in quantum and
coherent optics. Examples include studies on the effect of the
geometric phase on the coherent excitation and photoionization
of atoms driven by an intense laser field [19], on the photon
statistics of the output field in a degenerate parametric
amplifier [20], and on coherent pulse propagation [21].
Recently, it has been shown [22] that the geometric phase arising
in the propagation of a single-mode electromagnetic field
through a nonlinear medium is sensitive to the photon
statistics of the initial field. A measurement of the geometric
phase would thus be a way to obtain information on the photon
distribution of the field. Other practical applications in
optics include the construction of achromatic phase-shifters
[23] using the geometric phase for white light phase-stepping
interferometry in surface-profile studies [24,25]. It is
therefore clear that coherent optics is eminently suited for
a practical realization of the geometric phase in
various cases, thus providing a valuable probe to study
non-classical states of radiation [26].

With this motivation, it is therefore of interest to determine
the geometric phase for different kinds of coherent states. In
particular, there is a wide class of Hamiltonians whose eigenstates are 
generalized coherent states, and it is for this class that we
calculate the Berry phase. We shall be concerned with the Berry phase in 
its orginal setting
: the cyclic, adiabatic variation of parameters in a Hamiltonian with a 
discrete, non-degenerate spectrum. Berry's seminal work [11] established 
a well-known formula for the geometric phase $\gamma_n$ of the $n$ th level, 
as the line integral of a certain vector field over a closed contour in 
parameter space. Earlier works on Berry phases in the context of 
squeezed coherent states [27]  make direct use of this formula. 
In this paper, we adopt a more general approach. We show that for a Hamiltonian
system with equally-spaced levels, $\gamma_n$ is a linear function of $n$. 
Hence all the information on the Berry phases of the various eigenstates is
contained in the corresponding phases $\gamma_0$ and $\gamma_1$ of the 
ground state and the first excited state, respectively. In turn, this implies
that in the semi-classical limit, the anholonomy (the Hannay angle) is 
simply the difference between $\gamma_0$ and $\gamma_1$. This relationship 
simplifies enormously the computation of the semi-classical anholonomy, 
besides clarifying exactly why the latter vanishes in some
cases, although the 
corresponding Berry phase does not.

The plan of the paper is as follows. In Sec. 2 we derive the linear 
relationship mentioned above between $\gamma_n,\gamma_0$ and
$\gamma_1$.  In Sec. 3 we use this to obtain an explicit
expression  for $\gamma_n$ for the generalized harmonic
oscillator coherent states, and show how earlier results follow as special
cases. Finally, in Sec. 4 we extend the discussion to sets of multiphoton
coherent states built up from the squeezed vacuum ground state.
Section 5 contains some concluding remarks.

\section{General formula for $\gamma_n$ for equally-spaced levels}
We begin with the simple observation that the geometric phase is specific 
to the actual system under consideration, in the following sense : in a given
Hamiltonian, a clear identification must first be made of the actual dynamical
(or `fast') variables {\bf r} versus the adiabatic, externally varied 
`slow' variables {\bf R}. In this sense, the Hamiltonian $H_1 = \hbar\omega
(\ad{a}a+\frac{1}{2})$ (where {\bf r} comprises $a$ and $\ad{a}$, {\bf R} is
represented by the single parameter $\omega$, and $[a,\ad{a}] =
1$) is not identical,  {\em a priori}, to the Hamiltonian $H_2 = p^2/(2m) + \frac{1}{2}
m\omega^2q^2$ (where {\bf r} comprises $q$ and $p$, {\bf R} stands for
$m$ and $\omega$, and $[q,p] = i\hbar$). Of course, $H_2$ may be {\em re-written
} in the form $H_2 = \hbar\omega [\ad{a}({\bf R}) a({\bf R}) + \frac{1}{2}]$ by defining
the {\em parameter-dependent} operators $a({\bf R}) = (m\omega/2\hbar)^{1/2}q +i
(2m\omega\hbar)^{-1/2} p$ and its hermitian  conjugate $\ad{a}({\bf R})$. Their
commutator turns out to be [$a({\bf R}), \ad{a}({\bf R})$] = 1
for {\em all} 
{\bf R}, and it is this invariance of the operator algebra that makes it 
convenient to {\em analyse} the Hamiltonian $H_2$ using its {\em
representation} in terms of $a({\bf R})$ and $\ad{a}({\bf R})$. Our 
approach is essentially based on this property adapted to more 
general cases, as we shall see. We mention in passing that the distinction 
drawn above between different Hamiltonians (exemplified here by $H_1$ and $H_2$) 
is what is essentially responsible for the fact [28] that a canonical
transformation can convert (wholly or partly) a geometric phase into a 
dynamical phase, or vice versa.

Consider a Hamiltonian $H({\bf R})$ with equally-spaced, non-degenerate
eigenvalues, where {\bf R} denotes the set of `external' parameters to be 
varied adiabatically in some physical range. A form for $H({\bf R})$ which 
describes all the systems of interest to us is given (up to constants) by the 
hermitian operator
\be
H({\bf R}) &=& \ad{G} ({\bf R})\, X({\bf R})\, G({\bf R})
\ee
where $X({\bf R})$ is a positive-definite, hermitian operator, together
with the equal-time commutation relation
\begin{equation}
\left [ X({\bf R})G({\bf R})\,, \,\ad{G}({\bf R}) \right ] = 1
\end{equation}
on a suitable Hilbert space of states, for every {\bf R}. For the standard
oscillator, $G = a$ while $X$ is a multiple of the unit operator. In more
general instances, as in the case of Hamiltonians whose eigenstates are 
certain coherent states [8], $X$ may be a non-trivial function of $\ad{a}a$. 
Equation (2) leads to $[XG , H] = XG$ and $[\ad{G} , H] = -\ad{G}$ for 
every {\bf R}. It is then readily deduced that the spectrum of $H({\bf R})$ 
is the set of non-negative integers, i.e., there exists normalized eigenstates
\ktwo{n}{{\bf R}} such that
\be
H({\bf R})\, \ktwo{n}{{\bf R}} &=& n \,\ktwo{n}{{\bf R}}\,,\, (n = 0,1,2,\ldots).
\ee
Further, since $XG$ and $\ad{G}$ act as lowering and raising operators, 
respectively, we have
\begin{equation}
X({\bf R})\,G({\bf R})\, \ktwo{0}{{\bf R}} = 0, 
\end{equation}
\begin{equation}
X({\bf R})\,G({\bf R})\, \ktwo{n}{{\bf R}} = c_n \,\ktwo {n-1}{{\bf R}},
\end{equation}
and 
\begin{equation} 
\ad{G}({\bf R})\, \ktwo{n}{{\bf R}} = d_n \,\ktwo{n+1}{{\bf R}},
\end{equation}
where the time-independent constants $c_n$ and $d_n$ can be determined
if we know also the commutators $[G , \ad{G}]$ and $[G , X]$. We note in 
passing that Eq.(2) implies that $[G,\ad{G}X] = 1$ so that we could also 
have chosen $G$ as the lowering operator and $\ad{G} X$ as the corresponding
raising operator. However, we shall use the choice made earlier, as it is 
more convenient for the calculations to be presented in Sec. 4.

Let the parameter {\bf R} be varied adiabatically and cyclically with a 
time period $T$. Denoting by \ktwo{n}{{\bf R}} the $n$ th eigenstate at $t=0$
and by ${\ktwo{n}{{\bf R}}}_T$ the state to which it evolves at time $T$, 
we have
\be
\ktwo{n}{{\bf R}}_T &=& \exp \, \left [ i\,\gamma_n -\frac{i}{\hbar} \int_{0}^{T} E_n ({\bf R}(t))
\,dt \right ] \, \ktwo{n}{{\bf R}}
\ee
where $E_n({\bf R})$ is the corresponding 
 eigenvalue of the Hamiltonian $H({\bf R})$. 
Using Eq.(3), we have in the present instance
\be
\ktwo{n}{{\bf R}}_T &=& \exp \left [i\,\gamma_n
-\frac{inT}{\hbar} \right ]\,  \ktwo{n}{{\bf R}},
\ee
keeping in mind that ${\bf R}(0) = {\bf R}(T)$. The Berry phase $\gamma_n$ is 
given by [11]
\be 
\gamma_n &=& i \oint \langle n,{\bf R} | ( \nabla _{{\bf R}} \ktwo{n}{{\bf
R}}) \cdot d{\bf R}
\ee
where the integral runs over a closed contour in parameter space. 
Analogous to Eq.(8), we have also
\be
\ktwo{n-1}{{\bf R}}_T &=& \exp \left [i\,\gamma_{n-1}
-\frac{i(n-1)T}{\hbar} \right ]\,
 \ktwo{n-1}{{\bf R}}.
\ee
But $X({\bf R}) G({\bf R})$ is the lowering operator for {\em each} value of 
{\bf R}, so that at time $T$ we must have
\begin{equation}
(\,X({\bf R}) \,G({\bf R})\,)_T \,{\ktwo{n}{{\bf R}}}_T = c_n
\,{\ktwo{n-1}{{\bf R}}}_T \; ,
\end{equation}
where we have denoted by $(\,X({\bf R})\, G({\bf R})\,)_T$ the annihilation 
operator at time $T$. However, we now recall that $c_n$ does not depend 
on $t$, as it is determined by the equal-time commutators $[G,\ad{G}]$ and 
$[G,X]$. Substituting from Eqs.(8) and (10) for the kets in Eq.(11), we find 
that $(XG)_T$ must be given by
\begin{equation}
(\,X({\bf R})\, G({\bf R})\,)_T = X({\bf R})\, G({\bf R})\,
\exp\, \left [i\,(\gamma_{n-1} - \gamma
_n + \frac{T}{\hbar}) \right ].
\end{equation}
As this operator relation holds good for every $n$, it follows immediately that $
(\gamma_n - \gamma_{n-1})$ must be independent of $n$. In other words,
\begin{equation}
\gamma_n = \gamma_0 +n(\gamma_1 - \gamma_0),
\end{equation}
which is also consistent with the requirement that ${[(\,X({\bf R})
G({\bf R})\,)_T}]^n$  
acting on ${\ktwo{n}{{\bf R}}}_T$ yield the state ${\ktwo{0}{{\bf
R}}}_T$\,. We 
note that the formula obtained for $\;\gamma_n\;$ is only contingent on the 
existence of (i) an equally-spaced spectrum, and (ii) raising and lowering
operators connecting the eigenstates. The corresponding classical anholonomy is
the Hannay angle (the shift in the angle variable), for which the familiar 
semi-classical connection gives the formula [29] $\Delta \theta = - \partial 
\gamma_n / \partial n$. From Eq.(13), we have therefore
\begin{equation}
\Delta \theta = \gamma_0 - \gamma_1 .
\end{equation}

\section{Anholonomies for squeezed coherent states}
The following results are well known [29,30] : the Berry phase
$\gamma_n = 0$ for the linear harmonic oscillator with
Hamiltonian $p^2/(2m) + \frac{1}{2}m\omega^2 q^2$ or
$\hbar\omega(\ad{a}a+\frac{1}{2})$, under the variation of the parameters
$m$ and $
\omega$, but the generalized oscillator with a cross-term $(pq+qp)$ may have $
\gamma_n \neq 0$. Classically, the quadratic Hamiltonian $Ap^2+2Bpq+Cq^2$ has a
non-vanishing Hannay angle if and only if there is also a {\em
rotation} of the axes of the ellipse in the $(q,p)$ plane under
the adiabatic, cyclic variation of $A,B$ and $C$ : mere
translation of its centre and scaling of its axes lead to
$\Delta \theta = 0$. Turning to coherent states [27], for the
displaced oscillator with Hamiltonian
$\hbar\omega[(\ad{a}-\alpha^*)(a-\alpha)+1/2]$, whose ground
state is a coherent state, one finds $\gamma_n \neq 0$, but
$\Delta \theta =0$, under the adiabatic variation of the
complex parameter $\alpha$. However, if one considers {\em
squeezed} coherent states, $\Delta \theta \neq 0$ under the
adiabatic variation of the squeezing parameter $\beta$.

We now consider the general deformation of the oscillator Hamiltonian that
includes the foregoing as special cases. Let
\be
H' &=& \frac{p^2}{2m} + \frac{1}{2} m\omega^2q^2 \; = \; \hbar \omega (\ad{a}a
+\frac{1}{2})
\ee
(where the dependence of $a$ and $\ad{a}$ on the parameters $m$ and $\omega$ 
is implicit). Transforming $H'$ with the squeezing operator [31]
\be
S(\beta,\beta^*) &=& \exp \,\left (\frac{\beta \ad{a}^2 -
\beta^* a^2}{2} \right )\;\;(\beta \; \epsilon
 \; {\cal C})
\ee
and the displacement operator
\be
D(\alpha,\alpha^*) &=& \exp \,( \alpha \ad{a} - \alpha^* a),\;\;
(\alpha \; \epsilon \; {\cal C})
\ee
we have the Hamiltonian 
\be
H &=& D(\alpha,\alpha^*)\,S(\beta,\beta^*) \,H'\,\ad{S}(\beta,\beta^*)\,\ad{D}
(\alpha,\alpha^*)\;.
\ee
Let $\{\kone{n}\} (n=0,1,2,\ldots)$ denote the eigenstates of $H'$, and 
${\cal H}_0$ the Fock space spanned by these states. We note that both
$S(\beta,\beta^*)$ and $D(\alpha,\alpha^*)$ are unitary operators in
${\cal H}_0$. Comparing Eq.(18) with Eq.(1) we identify the operators
\be
G({\bf R}) &=& D\,S\,a\,\ad{S} \,\ad{D}\;, \;\;\; X({\bf R})\,=\,1
\ee
where
\begin{equation}
{\bf R} \; = \; \{m,\,\omega,\,\alpha = \alpha_1+i\alpha_2,\,\beta=\beta_1
+i\beta_2\}\;.
\end{equation}
Moreover
\begin{equation}
\left [ \,G({\bf R}),\ad{G}({\bf R})\, \right ] = 1
\end{equation}
in this case. Also
\be
H({\bf R})\, \ktwo{n}{{\bf R}} &=& \hbar\omega(n+\frac{1}{2}) \,\ktwo{n}{{\bf R}},
\ee
where
\begin{equation}
\ktwo{n}{{\bf R}} = D(\alpha,\alpha^*)\, S(\beta,\beta^*) \,\kone{n}\;,\; n=0,1,2,\ldots.
\end{equation}
Therefore, under an adiabatic, cyclic variation of the six real parameters 
comprising {\bf R}, the Berry phase and Hannay angle are given by Eqs.(13) and (14) 
respectively. Hence we have merely to compute explicitly the quantities
\begin{equation}
\gamma_0 = i \oint \, \langle 0,{\bf R}| \nabla_{\bf R}\ktwo{0}{{\bf R}} \cdot
d{\bf R}
\end{equation} 
and
\begin{equation}
\gamma_1 = i \oint \, \langle 1,{\bf R}| \nabla_{\bf R}\ktwo{1}{{\bf R}} \cdot
d{\bf R}\;.
\end{equation} 
(The gradient is understood to act on the ket to its right.) Now
$\langle 1,{\bf R}| \nabla_{\bf R}\ktwo{1}{{\bf R}}$ can be
simplified by noting that $\ktwo{1}{{\bf R}} = \ad{G}({\bf R})\,
\ktwo{0}{{\bf R}} ,\;\langle 1,{\bf R}| =\,
\langle 0, {\bf R}|\, G({\bf R})$. Moreover, using the fact that $[\,G({\bf R}),
\ad{G}({\bf R})\,] = 1$ in this case, and that $G({\bf R})
\,\ktwo{0}{{\bf R}} =0$, we find
\begin{equation}
\langle 1,{\bf R}| \nabla_{{\bf R}} \ktwo{1}{{\bf R}}  = 
\langle 0,{\bf R}| \nabla_{{\bf R}} \ktwo{0}{{\bf R}}  + 
\langle 0,{\bf R}|\, \left [\,G({\bf R}),\left (\nabla_{{\bf R}}
\ad{G}({\bf R})\right )\,\right ]
\,\ktwo{0}{{\bf R}}\;.
\end{equation}
Therefore
\begin{equation}
\gamma_1 = \gamma_0 +i\oint \,\langle 0,{\bf R}|\,\left [G({\bf
R}),\left (\nabla_{{\bf R}}
 \ad{G}({\bf R})\right )\,\right ] \ktwo{0}{{\bf R}} \cdot d{\bf R}\;.
\end{equation}
The emergence of the first term $(\gamma_0)$ on the right-hand side is 
entirely a consequence of the commutation relation $[G,\ad{G}] = 1.\,$(In Sec.$\,$4,
we shall see what happens when $X \neq 1$, \mbox{$[G,\ad{G}] \neq 1$)}. To evaluate
the commutator in Eq.(27), it is helpful to use the fact that Eq.(19) can be 
reduced to the explicit expression [16]
\begin{equation}
G({\bf R}) = (a-\alpha) \cosh |\beta| - (\ad{a}-\alpha^*) \frac{\beta}{|\beta|}
\sinh |\beta|.
\end{equation}
Carrying out the calculations involved (the salient features are given in the 
Appendix), we arrive finally at the following results. It turns out that 
variations in $m$ and $\omega$ are both included in that of the single parameter
\begin{equation}
\lambda = \mbox{ln} (m\omega).
\end{equation}
Moreover, there occurs a natural separation of the contributions of the 
squeezing and displacement parameters to the Berry phase $\gamma_n$ acquired
by \ktwo{n}{{\bf R}}. We find
\begin{equation}
\gamma_n = {\gamma_n}^{(D)} + {\gamma_n}^{(S)}
\end{equation}
with
\begin{equation}
{\gamma_n}^{(D)} = \oint \,(\alpha_2 \,d\alpha_1 -\alpha_1 \,d\alpha_2 -\alpha_1
\alpha_2 \,d\lambda)
\end{equation}
and 
\begin{equation}
{\gamma_n}^{(S)} = (n+\frac{1}{2}) \oint \left ( \frac{\sinh |\beta|}{|\beta|}
\right )^2 (\beta_2 d\beta_1 - \beta_1 d\beta_2) - \frac{\beta_2}{|\beta|}
\,\sinh |\beta| \,\cosh |\beta| d\lambda \; ,
\end{equation}
where $\oint$ stands for the integral over the closed contour traversed in the 
space of the six parameters {\bf R}. We are now ready to read off a number of
special cases.
\\
(i) $\alpha$= {\em constant}, $\beta$= {\em constant}, $m$ {\em
and} $\omega$  {\em varied} : It  is 
evident that $\gamma_n= 0$, and hence $\Delta \theta = 0$, in this case.
Varying $m$ and $\omega$ does not produce a geometric phase, as the 
variation appears as a perfect differential, $d\,(\mbox{ln}\, m\omega)$. The 
original oscillator corresponds to the trivial case $\alpha = 0, 
\beta = 0$.
\\
(ii) $\beta = constant$ : In this case (which includes $\beta= 0$, or no 
squeezing) we have a non-vanishing Berry phase which is just $\gamma_n 
= \gamma_n^{(D)}$, but this is $n$-independent, so that the Hannay angle 
$\Delta \theta = 0$. This remains so, of course, even if $m$ and
$\omega$ are also kept constant, and $(\alpha_1, \alpha_2)$ alone are 
varied, as found in Ref. [27]. Writing $\gamma_n$ as the line integral of a 
vector potential [11], it is evident that this latter case ($\lambda = $
constant) implies a vector potential ${\bf A}$ with components $(\alpha_2,-\alpha_1,
0)$ along the $\alpha_1, \alpha_2$ and $\lambda$ directions. The corresponding
``magnetic field'' ${\bf V}= \nabla_{{\bf R}} \times {\bf A}$ is therefore a {\em uniform}
field along the $\lambda$-direction; the Berry phase is thus equal, in 
magnitude, to twice the area enclosed by the loop in the ($\alpha_1,\alpha_2$)
plane. On the other hand if $\lambda$ is {\em also} varied along with 
$\alpha_1$ and $\alpha_2$, the vector potential ${\bf A} = (\alpha_2, -\alpha_1
-\alpha_1\alpha_2)$. It is interesting to note how the variation in $\lambda$ 
gets coupled to the displacement parameters $\alpha_1$ and $\alpha_2$. The field 
${\bf V}$ now involves a singular source over and above the earlier uniform field : a line 
singularity (``anti-vortex'') along the $\lambda$-axis, with winding number 
equal to -1.
\\
(iii)$\alpha = constant$ : In this case (which includes $\alpha=0$, or no 
displacement) we have an $n$-dependent Berry phase, and therefore a non-zero
\, $\Delta \theta$. This remains true if $\lambda$ is also kept constant and only 
$\beta_1,\beta_2$ are varied [27]. Then
\begin{equation}
\gamma_n = - \left ( n+\frac{1}{2} \right ) \oint \sinh^2 |\beta|\, d\, (\arg \beta)
\end{equation}
corresponding to a magnetic field normal to the $\beta$-plane of 
magnitude $(n+\frac{1}{2}) \sinh (2|\beta|)/|\beta|$. We note also that a non-vanishing 
$\gamma_n$ occurs if $\beta_1$ and $\lambda$ alone are varied, provided the {\em 
imaginary} part $\beta_2$ of the squeezing parameter $\beta$ is non-zero. In 
this connection, it is useful to note that the Hamiltonian $\hbar\omega \,S(\beta,
\beta^*)\, (\ad{a}a+1/2)\,\ad{S} (\beta, \beta^*)$ corresponding to  pure 
squeezing can be written, in terms of the original oscillator operators $q$ and 
$p$, as $Ap^2+B(pq+qp)+Cq^2$, with
\be
A &=& \frac{1}{2m} [\cosh^2 |\beta| +\sinh^2 |\beta| +2 \frac{\beta_1}{|\beta|}
\cosh|\beta| \sinh|\beta|], \\
B &=& -\frac{\beta_2}{|\beta|} \cosh|\beta| \sinh|\beta|, \\
C &=& \frac{m\omega^2}{2} [\cosh^2 |\beta| +\sinh^2 |\beta| -2 \frac{\beta_1}{|\beta|}
\cosh|\beta| \sinh|\beta|].
\ee
\section{Anholonomies for multiphoton squeezed coherent states}
We turn now to the application of our formalism to Hamiltonians based on 
multiphoton coherent states. To be specific, we consider the eigenstates 
of the square of the annihilation operator. We begin with the observation [8] that the 
commutation relation
\begin{equation}
\left [ \frac{1}{2} (1+\ad{a}a)^{-1} a^2 , \ad{a}^2 \right ] = 1
\end{equation} 
is valid on the {\em even} subspace ${\cal H}_1 =
\{ span \kone{2n}; n=0,1,
\ldots\}$ of ${\cal H}_0$. (It is in fact valid on ${\cal H}_0 -
span \kone{1}$, but for our present purposes we restrict
attention to ${\cal H}_1$). Comparing Eq.(37) with Eq.(2), we
identify the raising and lowering operators $\ad{G}$ and $XG$
according to
\begin{equation}
\ad{G} = \ad{a}^2 \;,\; X=\frac{1}{2}(1+\ad{a}a)^{-1}\;.
\end{equation}
The `Hamiltonian' $\ad{G}XG$ itself is easily verified to have
matrix elements identical to those of $\ad{a}a/2$, as one might
have anticipated. However, it is not this Hamiltonian in which
we are interested, but rather in the anholonomies associated
with its deformations which have generalized coherent and/or
squeezed states (eigenstates of $XG$) as their ground states.

We therefore define the corresponding displacement operator
\begin{equation}
D(\alpha,\alpha^*) = \exp \,(\alpha \,\ad{G} - \alpha^* \,XG)\;.
\end{equation}
The state $D\,\kone{0}$ is then an eigenstate of $XG$ with eigenvalue $\alpha$. 
The next step is to attempt to construct a Hamiltonian $D\,(\ad{G}XG)\,D^{-1}$ whose 
ground state would be the coherent state $D\,\kone{0}$ (rather than the vacuum \kone{0})
, so that we may proceed as in Sec. 3 to investigate the associated anholonomies. 
Unfortunately, the displacement operator in Eq.(39) is no longer unitary, 
so that $D\,\ad{G}XG\,D^{-1}$ is not hermitian. It is evident that the problem arises
because the raising operator $\ad{G} = \ad{a}^2$ and its conjugate $a^2$ do not 
satisfy the commutation relation $[G,\ad{G}] =1$; rather, it is the commutator 
$[XG,\ad{G}]$ that is equal to unity. One way out is to make a different identification
of $G$ and$X$ than that made in Eq.(38), and we shall consider this possibility 
subsequently. For the present, we note that there is another approach, based on 
squeezing rather than displacement:

The squeezing operator $S(\beta,\beta^*)$ defined in Eq.(16) can be expanded 
[31] in the normal-ordered form 
\be
S(\beta,\beta^*) &=& (\cosh |\beta|)^{-\frac{1}{2}} \exp \,\left
( \frac{\ad{a}^2 \beta}{2|\beta|}
\tanh |\beta| \right ) \nonumber \\
 && \left ( \sum_{r=0}^{\infty} \frac{(\mbox{sech} |\beta|-1)^r}{r!} \ad{a}^r a^r
\right ) \exp \,\left ( -\frac{a^2 \beta^*}{2|\beta|}  \tanh
|\beta| \right ).
\ee
With the help of this expansion, we may establish that 
\begin{equation}
\frac{1}{2} (1+\ad{a}a)^{-1}\, a^2 \,S(\beta,\beta^*)\,\kone{0} = 
\left[ \frac{\beta}{2|\beta|} \tanh |\beta| \right ] S(\beta,\beta^*)\kone{0}.
\end{equation}
In other words, the squeezed vacuum 
\begin{equation}
\ktwo{0}{\beta} \equiv S(\beta,\beta^*)\,\kone{0}
\end{equation}
is also a generalized coherent state (an eigenstate of the lowering operator 
$XG$). Moreover, $S$ is unitary. We may therefore construct the deformed, 
hermitian Hamiltonian (restoring the appropriate constants)
\begin{equation}
H_S = \frac{1}{2} \hbar \omega \,S(\beta,\beta^*)\, \left [\ad{a}^2 (1+\ad{a}a)^{-1}\,
a^2 +1 \right ]\, \ad{S} (\beta,\beta^*)\;.
\end{equation}
The ground state of this Hamiltonian is the state
\ktwo{0}{\beta} defined in  
Eq.(42) :
\begin{equation}
H_S \,\ktwo{0}{\beta} = \frac{1}{2} \hbar\omega \,\ktwo{0}{\beta}.
\end{equation}
The raising and lowering operators for this system are 
\begin{equation}
\ad{G}({\bf R}) = S(\beta,\beta^*)\, \ad{a}^2 \ad{S}(\beta,\beta^*)
\end{equation}
and
\begin{equation}
X({\bf R})\, G({\bf R}) = \frac{1}{2} S(\beta,\beta^*)\,(1+\ad{a}a)^{-1} a^2\,
\ad{S}(\beta,\beta^*)
\end{equation}
respectively. The excited state \ktwo{2n}{\beta} ($n = 1,2,\ldots)$ of 
$H_S$ is obtained by applying $(\ad{G}({\bf R}))^n$ to \ktwo{0}{\beta},
and the corresponding eigenvalue is $\hbar\omega(n+1/2)$.

We may now consider the Berry phase acquired by the state \ktwo{2n}{\beta}
under the adiabatic, cyclic variation of the four parameters $m, \omega,
\beta_1$ and $\beta_2$ in $H_S$ (the first two being implicitly contained in 
$a$ and \ad{a} as before). The answer, in fact, may be written down directly 
from our earlier results once we recognise that \ktwo{2n}{\beta} is {\em also}
given by
\begin{equation}
\ktwo{2n}{\beta} = S(\beta,\beta^*)\, \kone{2n}\;,
\end{equation}
i.e., raising with $(\ad{G})^n$ and squeezing with $S$ can be performed in either
order. The Berry phase is therefore precisely $\gamma_{2n}^{(S)}$, where 
$\gamma_n^{(S)}$ is given by Eq.(32), and the rest of the discussion proceeds 
as before. 

Finally, let us return to the Hamiltonian (or number operator)
\begin{equation}
N' = \frac{1}{2} \ad{a}a = \frac{1}{2} \,\ad{a}^2 (1+\ad{a}a)^{-1} a^2
\end{equation}
which, in the space ${\cal H}_1$, has eigenvalues 0,1,2, .... The question is 
whether we can write $N'$ in the form $\ad{a_1}a_1$ where $\ad{a_1}$ and $a_1$ are
the corresponding raising and lowering operators and, {\em moreover}, $
[a_1,\ad{a_1}] = 1$ in ${\cal H}_1$. This would avoid the problem encountered
earlier which arose because $[G,\ad{G}]$ was not equal to the unit operator. 
Now, since $(1+\ad{a}a)^{-1}$ is a bounded positive operator, there exists 
(according to the square-root lemma [32]) a unique positive bounded operator
$(1+\ad{a}a)^{-\frac{1}{2}}$ whose square is $(1+\ad{a}a)^{-1}$. We 
may therefore write
\begin{equation}
N' = \ad{a_1} a_1
\end{equation}
with
\begin{equation}
a_1 = 2^{-\frac{1}{2}} (1+\ad{a}a)^{-\frac{1}{2}} a^2 \;, \ad{a_1} =
2^{-\frac{1}{2}} \ad{a}^2 (1+\ad{a}a)^{-\frac{1}{2}}\;,
\end{equation}
and $[a_1, \ad{a_1}] = 1$ in ${\cal H}_1$. (It is clear [33] that 
$(a_1, {\cal H}_1)$ is isomorphic to $(a,{\cal H}_0)$, each of these constituting
an irreducible representation of the basic commutation relation $[F,\ad{F}] =1$). 
The procedure followed in Sec. 3 for the original oscillator Hamiltonian may now be
repeated, unaltered : {\em unitary} displacement and squeezing operators 
\begin{equation}
D_1 (\alpha,\alpha^*) = \exp \,(\alpha\ad{a_1} - \alpha^* a_1)
\end{equation}
and
\begin{equation}
S_1 (\beta,\beta^*) = \exp \,\left ( \frac{\beta\ad{a_1}^2 - \beta^* a^2_1}{2}
\right )
\end{equation}
may be used to deform $N'$, corresponding squeezed coherent states constructed, 
and their anholonomies derived, exactly as in Sec. 3. It is also evident that
the same process can be repeated in the (isomorphic) subspaces
${\cal H}_2 \supset
{\cal H}_3  \supset \ldots ,$ where ${\cal H}_k = \{ span
\,\kone{2^k n} \}$, by defining the
operators $a_k,\ad{a_k} $ in ${\cal H}_k$  recursively, according to
\begin{equation}
a_k = 2^{-\frac{1}{2}} (1+\ad{a}_{k-1} a_{k-1})^{-\frac{1}{2}} \,a_{k-1}^2\;\;,
\end{equation}
so that $[a_k,\ad{a_k}] = 1$ in ${\cal H}_k$.

\section{Conclusions}

In this paper we have shown that, for an arbitrary Hamiltonian
with equally-spaced, non-degenerate eigenvalues, the geometric
phase $\gamma_n$ of the n th eigenstate is a linear function of
$n$. Crucial to the derivation of this result is the existence
of raising and lowering operators (connecting the different
states,) that satisfy a definite algebra. Using the above
formalism, the geometric phase was calculated both for
generalized squeezed coherent states and for a class of
multi-photon coherent states.

A natural question that arises is whether our approach can be
extended to the case of Hamiltonians with unequally-spaced
levels. Although the existence (and construction) of appropriate
raising and lowering operators is not immediately obvious in the
general case, one possible avenue of approach is the
factorization method [34] and its recent extensions,
particularly in the context of supersymmetric quantum mechanics
[35]. This question is presently under investigation.

Finally, it turns out to be possible to construct coherent
states (and generalized coherent states) for a class of
Hamiltonians which are strictly isospectral to the harmonic
oscillator [36]. While certain classes of these states are
essentially unitarily equivalent to those obtained from the
original oscillator Hamiltonian, other classes of coherent
states can be constructed, via supersymmetry transformations,
that are not unitarily equivalent to the original ones. The
geometric phases associated with such states are also under
investigation, and the results will be reported elsewhere. 

\vspace{.2cm}

\noindent {\bf Acknowledgments}\\
We thank S. Chaturvedi, M.V. Satyanarayana, M.D. Srinivas and V. Srinivasan for useful 
discussions, and a referee for helpful comments.

\newpage
{\bf Appendix}\\
\renewcommand{\theequation}{A.\arabic{equation}}
We outline the steps leading to the explicit formulas given in
Eqs.(30)-(32) for the Berry phase $\gamma_n$ corresponding to
the squeezed, displaced oscillator Hamiltonian $H$ of Eq.(18).

To evaluate $\gamma_0$, given by Eq.(24), we work in the
position representation, in which
\setcounter{equation}{0}
\begin{equation}
\gamma_0 = i\, \oint \, d{\bf R} \cdot \left [ \, \int \, dx \,
{\psi_0}^*\, (x;{\bf R}) \, \nabla_{{\bf R}}\,\psi_0 \, (x;{\bf R})
\,\right ] 
\end{equation}
where {\bf R} stands for the set of six parameters $\{
m,\,\omega,\,\alpha=\alpha_1+i\alpha_2,\,\beta=\beta_1+i\beta_2\}$.
The ground state wavefunction $\psi_0$ is found in a
straight-forward  manner, and is
given by
\begin{equation}
\psi_0 \, (x;{\bf R}) = {\left
(\frac{m\omega}{2\pi\hbar} \right )}^{\frac{1}{4}}\, 
\exp\,\left( -\frac{v_1^2}{4u_1} \right )\,\exp\,\left
(-\frac{1}{2} ux^2 - vx \right )\;, 
\end{equation}
with
\begin{equation}
u = u_1 +i u_2 = \left ( \frac{m\omega}{2\hbar} \right ) \left (
\frac{|\beta|-i\beta_2 \, \sinh 2|\beta|}{|\beta| \cosh 2|\beta|
+ \beta_1 \sinh 2|\beta|} \right )\;, 
\end{equation}

\begin{equation}
v = v_1 +i v_2 = \left ( \frac{2m\omega}{\hbar} \right )^{\frac{1}{2}}
\left (\frac{-\alpha_1|\beta| + i (\alpha_1\beta_2 -
\alpha_2\beta_1) \sinh 2|\beta| - i \alpha_2 |\beta| \cosh
2|\beta|}{|\beta|\cosh2|\beta|+\beta_1 \sinh 2|\beta|} \right )\;.
\end{equation}
\\ 
Next, we calculate the partial
derivatives of $\psi_0$ with respect to the six parameters (it
is convenient to consider the logarithmic derivative of
$\psi_0$), substitute these in Eq.(A.1) and carry out the
(Gaussian) integrals over $x$, to arrive at the result
\be
\gamma_0 &=& \oint \,\left [(\alpha_2 \,d\alpha_1 -\alpha_1 \,d\alpha_2 -\alpha_1
\alpha_2 \,d\lambda) +  \frac{1}{2} \left ( \frac{\sinh |\beta|}{|\beta|}
\right )^2 (\beta_2 d\beta_1 - \beta_1 d\beta_2)  \right.
\nonumber \\ 
&& \left. - \frac{1}{2}\frac{\beta_2}{|\beta|}
\,\sinh |\beta| \,\cosh |\beta| d\lambda \right ]\;,
\ee
where $\lambda = \mbox{ln}\, (m\omega)$ as defined in Eq.(29).

We must now compute $\gamma_1$ from Eq.(27). Using the
representation given in Eq.(28) for the operator $G({\bf R})$
(and remembering that  $m$ and $\omega$ occur in the
expressions for $a$ and \ad{a}), we find 
\begin{equation}
\left [ \,G({\bf R}), (\nabla_{{\bf R}} \ad{G}({\bf R}) \,\right ] \cdot
d\,{\bf R} =  i \,\frac{\beta_2}{|\beta|}
\,\sinh |\beta| \,\cosh |\beta| d\lambda   
  -i  \left ( \frac{\sinh |\beta|}{|\beta|}
\right )^2 (\beta_2 d\beta_1 - \beta_1 d\beta_2).
\end{equation}

There is no operator dependence left in this expression because
$[a,\ad{a}] = 1$. Moreover, since $\psi_0 \, (x;{\bf R})$ is
normalized to unity, Eq.(27) becomes
\begin{equation}
\gamma_1 = \gamma_0 +\oint \, \left [ \left ( \frac{\sinh |\beta|}{|\beta|}
\right )^2 (\beta_2 d\beta_1 - \beta_1 d\beta_2) - \frac{\beta_2}{|\beta|}
\,\sinh |\beta| \,\cosh |\beta| d\lambda \right ]\;.
\end{equation}
 
Substitution of Eqs.(A.5) and (A.7) in the general formula
for $\gamma_n$ (Eq.(13)) yields the results quoted in Eqs.(30)-(32).

\end{document}